\documentclass[12pt]{iopart}

\usepackage{iopams}  
\usepackage{color}
\usepackage{soul}
   \usepackage[pdftex]{graphicx}

\begin{document}

\title[Side-wall spacer passivated sub-$\mu$m Josephson junction fabrication process]{Side-wall spacer passivated sub-$\mu$m Josephson junction fabrication process}

\author{Leif Gr\"{o}nberg, Mikko Kiviranta, Visa Vesterinen, Janne Lehtinen, Slawomir Simbierowicz, Juho Luomahaara, Mika Prunnila, and Juha Hassel}

\address{VTT Technical Research Centre of Finland, Tietotie 3, FI-02150 Espoo, Finland}
\ead{Juha.Hassel@vtt.fi}
\vspace{10pt}
\begin{indented}
\item[]September 2017
\end{indented}

\begin{abstract}
We present a structure and a fabrication method for superconducting
tunnel junctions down to the dimensions of 200 nm using i-line UV lithography. The
key element is a side-wall-passivating spacer structure (SWAPS) which is shaped for smooth crossline
contacting and low parasitic capacitance. The SWAPS structure enables
formation of junctions with dimensions at or below
the lithography-limited linewidth. An additional benefit is avoiding the excessive use
of amorphous dielectric materials which is favorable in
sub-Kelvin microwave applications often plagued by nonlinear and lossy dielectrics. 
We apply the structure to niobium trilayer junctions,
and provide characterization results yielding evidence on wafer-scale scalability, and
critical current density tuning in the range of 0.1 -- 3.0 kA/cm$^2$. We discuss the
applicability of the junction process in the context of different applications, such as,  SQUID magnetometers and
Josephson parametric amplifiers.
\end{abstract}

\section{Introduction}
Superconducting tunnel junctions form the basis of many applications in the field of low-temperature sensors and electronics. A multitude of different process versions has been introduced in the past decades optimized for different applications \cite{hyp1,aist1,jen1,mul1}. The starting point of this work is our established fabrication line  based on niobium trilayer junctions  \cite{gro1,kiv1} that has been extensively utilized in different applications and academic explorations  such as biomagnetism \cite{vesa1}, astronomical imaging applications \cite{kiv2}, Josephson microwave amplifiers \cite{ ves1}, and demonstrators of parametric quantum effects \cite{lah3}, to mention a few. In this paper we present a modified scheme for junction patterning based on a smooth side-wall-passivating spacer structure (SWAPS) enabling an accurate definition of the junction by a self-aligned cross-type structure. Similar spacers have been previously used in other applications \cite{fra1}. To our knowledge such spacers have not been used in the context of superconducting junctions while we have used a similar technique to assist step-coverage in superconductive cross-overs \cite{kiv1} (marked proprietary). Cross-type superconducting junctions have been previously introduced by others  \cite{jen1,aoy1,dan1}: a common aspect is that such processes enable, for a given lithography resolution, minimal junction size and very small parasitic capacitance. For our process, we present characterization data verifying that junctions with dimensions down to 0.2 $\mu$m $\times$ 0.2 $\mu$m can be reliably produced with the process based on 150 mm wafers and UV optical lithography. 

In addition to minimizing the junction size and parasitic capacitance, our approach aims at solving another issue of significance especially in sub-Kelvin applications utilizing microwave-resonant structures. It is a well-known feature of common amorphous dielectric materials that they possess excess microwave loss as well as power and temperature nonlinearity attributed to the two-level states (TLSs) present in the material \cite{mar1}.  This is typically considered in the context of decoherence mechanisms in quantum bits. We have found that the effects are also harmful within Josephson parametric amplifiers (JPAs),  especially thanks to the temperature dependence of the real part of the dielectric constant \cite{ves1}. These problems are not strictly due to the capacitances of the junctions themselves, but relate to the parasitic effect caused by  the dielectric material that overall couples to the microwave resonant features of the device. For this reason, in our junction definition scheme unlike in cross-layer processes based on planarization,  the dielectric materials are removed from the device area excluding the junction sidewall itself.

\section{Fabrication techniques and samples}

\begin{figure}[!t]
\begin{center}
\includegraphics[width=15cm]{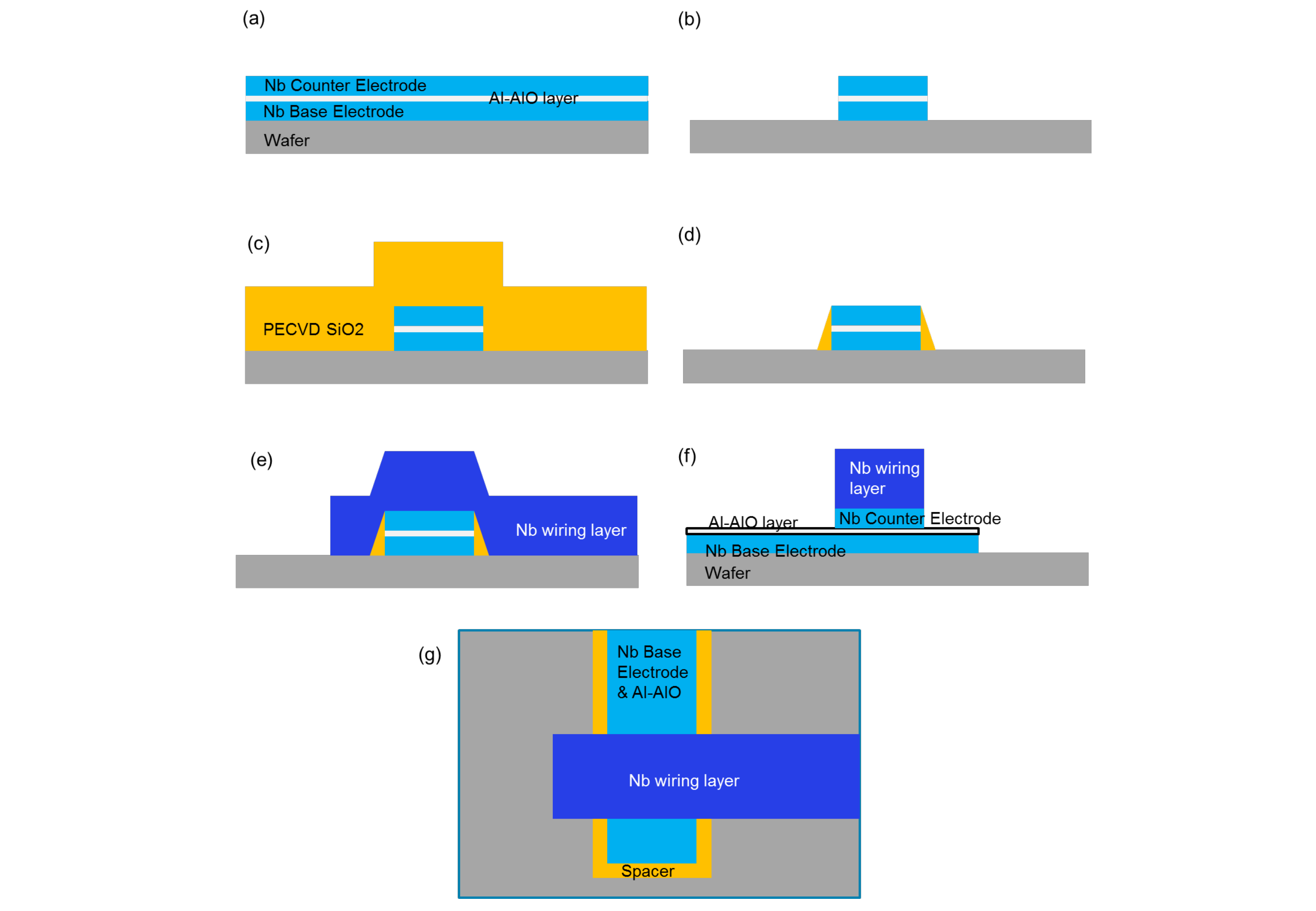}
\end{center}
 \caption{Illustration of the main process steps in the junction formation. (a) Trilayer deposited and oxidized. (b) Trilayer patterned. (c) Spacer material
(PECVD SiO$_2$) deposited. (d) Spacer formed after blanko SiO$_2$ etch. (e,f) Wiring layer deposited and patterned. The view in (f) is 90 degrees rotated. (g) Top view of the final junction structure. The junction is formed at the area of the crossing. } \label{prosteps}
\end{figure}

\begin{figure}[!bt]
\begin{center}
\includegraphics[width=14cm]{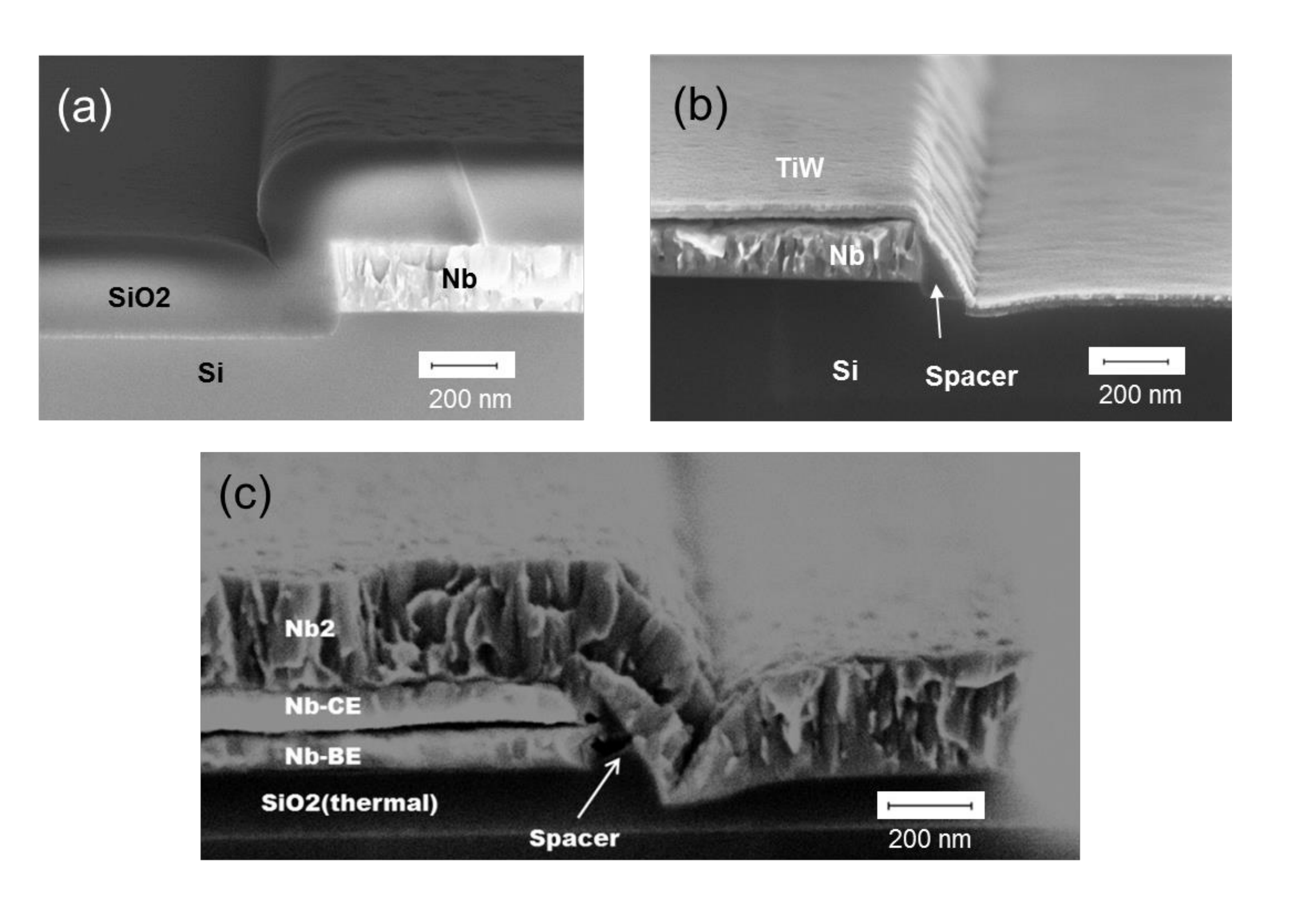}
\end{center}
 \caption{Scanning electron micrographs of SWAPS process test structures: a) A test structure with PECVD SiO$_2$ deposited to cover the Nb edge. b) A similar test structure after the spacer-forming etch step has been executed. Titanium tungsten (TiW) is used for contrast enhancement only. c) The complete Nb trilayer junction stack. The structures of frames a) and b) are formed on plain silicon substrate, while the structure of frame c) is on thermally oxidized silicon. }\label{SEM}
\label{System}
\end{figure}

The process steps are illustrated in Fig. \ref{prosteps}, and scanning electron micrographs of cross sections of selected intermediate and finished spacer structures are shown in Fig. \ref{SEM}. The basis of the structure is the trilayer  (Fig. \ref{prosteps}(a))  that is formed in a standard fashion by sputtering the base Nb layer, and on top of that a thin ($\sim$10 nm) Al layer \cite{tri1}. Aluminium surface is then oxidized by letting oxygen into the chamber. The tunnel resistance/critical current density is determined by the oxygen exposure \cite{exposure1,exposure2}. The counter electrode Nb is sputtered on top of AlO$_x$. This is all done in situ without breaking the vacuum.  The thicknesses of both base and counter Nb electrodes are 100 nm. The trilayer is patterned by i-line UV projection lithography with nominal linewidth resolution of 350 nm, and plasma etched to form a strip (Fig. \ref{prosteps}(b)). The etch is performed in one etching sequence containing three sequential steps removing the counter electrode, the Al/AlO$_x$ layer, and the base electrode, respectively.  For the Nb layers the reactive plasma is a mixture of Cl$_2$, CF$_4$ and Ar. For Al/AlO$_x$ removal the plasma contains BCl$_3$, Cl$_2$ and Ar. A layer of plasma enhanced chemical vapor deposited (PECVD) SiO$_2$ is grown in a parallel electrode reactor (PECVD Oxford Plasmalab 100) to a thickness of the step height in a standard way from SiH$_4$ and N$_2$O (Fig. 1(c))\cite{ada1}. Figure \ref{SEM}(a) illustrates the realized step coverage with some cusping \cite{fra1} present at the corner.  The spacer structure (Fig. \ref{prosteps}(d)) is formed when performing anistropic plasma back etching of the PECVD SiO$_2$ layer.  Anisotropy is achieved with CHF$_3$ in the plasma which produces a polymer that passivates the side wall  during etching \cite{ste1}. The realized spacer structure is shown in Fig. \ref{SEM}(b). 
This etch step could be replaced with techiques such as chemical-mechanical planarization \cite{ket1}, or liftoff-based planarization \cite{jen1}, the drawback of which as compared to SWAPS is that the dielectrics are left  essentially everywhere in the device area.  Finally, the wiring Nb layer is deposited on top, patterned by projection lithography, and plasma etched to strip geometry (Fig. \ref{prosteps}(e,f)).  The top view of the cross junction is shown in Fig. \ref{prosteps}(g). A scanning electron micrograph of a realized junction stack is shown in Fig. \ref{SEM}(c).

We note that on top of the junction scheme described above we have up to date developed process versions for two applications, namely SQUID magnetometers and
Josephson parametric amplifiers. The versions differ from each other in terms of substrate quality (thermally oxidized/non-oxidized, respectively), and in terms of additional layers for superconducting crossings, flux input coupling, and resistive shunts. While the main emphasis here is to verify the functionality of the junction process, the data shown below are from three different wafers fabricated either just for junction testing (wafer A) or for SQUID magnetometry (wafers B and C). For wafers B and C we thus have data from resistively shunted SQUID structures as well. From the junction point of view the wafers differ from each other mainly in terms of oxygen exposure $E$ used in the tunnel barrier formation. For wafers A,B, and C it applies $E\approx$ 1.3 kPa-s, 3.5 kPa-s, and 1600 kPa-s, respectively.

\section{Electrical characteristics}

\begin{figure}[t!]
\begin{center}
\includegraphics[width=17cm]{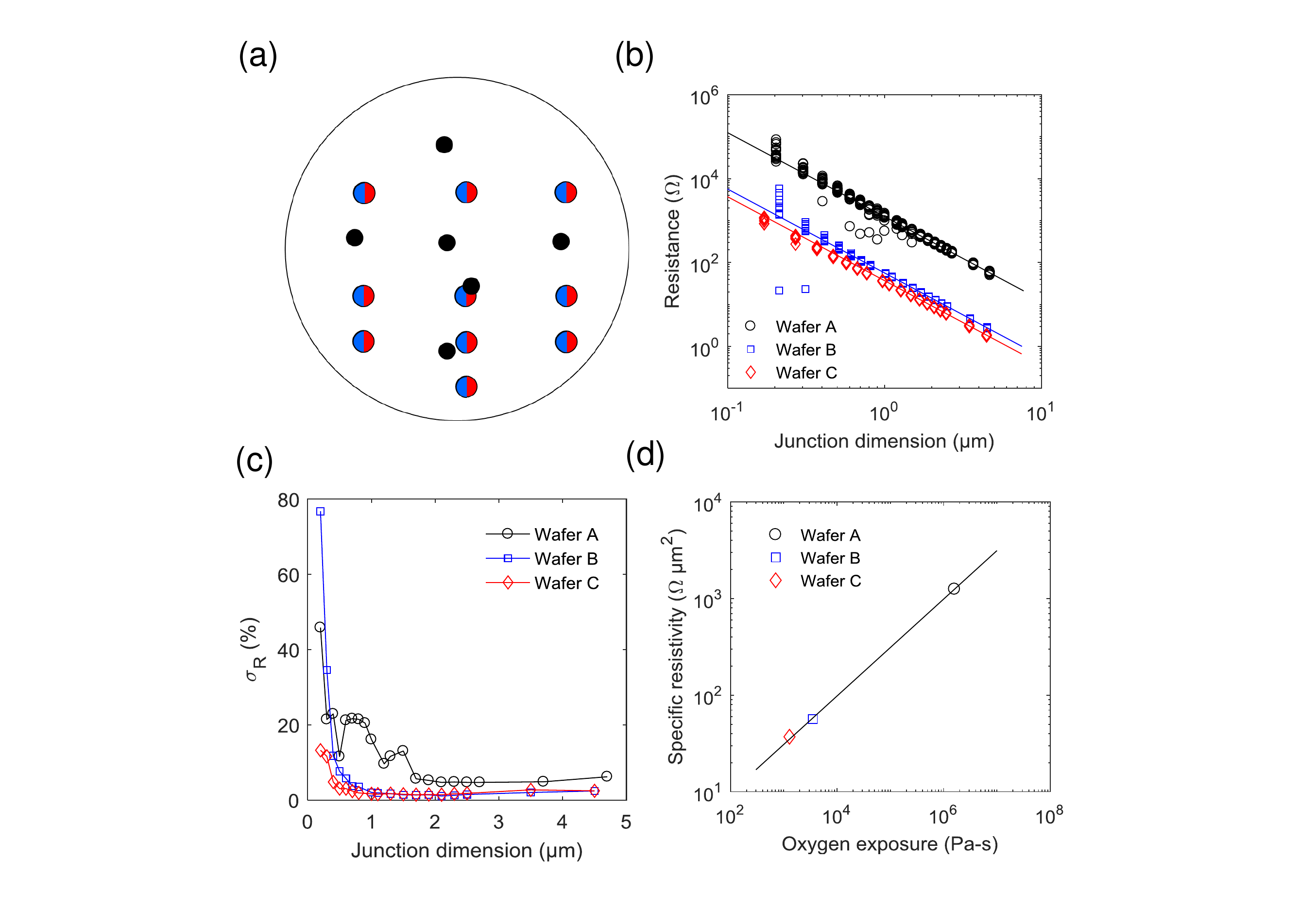}
\end{center}
 \caption{(a) The locations of the chips on 150 mm wafer from which the room temperature resistance data was measured: black small circles correspond to the chips of wafer A and red-blue large circles the locations on wafers B and C. (b) Junction resistance as the function of junction dimension, i.e. the side length of the square shaped junctions. (c) Standard deviation $\sigma_\mathrm{R}$ for resistances across the wafer normalized to the median resistance of each junction size. (d) Specific tunnel junction resistivity plotted as the function of oxygen exposure. The three experimental values are obtained as fitting parameters from the data of (b), and the solid line is the power-law fit.}\label{resmeas}
\label{System}
\end{figure}

Room-temperature tunnel resistance data from the three wafers is presented in Fig. \ref{resmeas}. For each wafer, the resistance measurement is performed on chips with on-wafer locations as  indicated in Fig. \ref{resmeas}(a). From each chip on wafer A we have measured junctions with 20 different sizes and 3 junctions per size, i.e. altogether 60 junctions per chip and 360 junctions per wafer. From wafers B and C we have measured junctions with 18 different sizes per chip and 1 junction per size, i.e., altogether 162 junctions per wafer. The resistance data as the function of the junction dimension are presented in Fig. \ref{resmeas}(b). The estimated junction sizes including the effects of the lithography and overetching for each wafer were obtained by fitting the tunnel conductance data with two free parameters, namely the junction specific resistivity and the size reduction as compared to the mask.  For wafers A,B and C we obtained the fitted size reductions of (0.29$\pm$0.02) $\mu$m, (0.52$\pm$0.01) $\mu$m, and (0.49$\pm$0.02) $\mu$m, respectively. The data of Fig. \ref{resmeas}(b) are already plotted as the function of the expected real size by subtracting this wafer-level size reduction from the junction dimensions. The difference between wafer A, and wafers B and C, may be related to the difference in the wiring layer thickness and related overetching time which was a factor of $\sim$2 longer for B and C in comparison to A. The similarity between B and C indicates good reproducibility from wafer to wafer when the etching parameters are held constant. The error bars represent the statistical standard error from the least squares fit.  

It has been previously observed that room temperature resistance data well predicts the low-temperature characteristics  \cite{gro1}. Thus it is useful to look at the statistics of the resistances that should give an indication on the expected critical current values and spread. Figure \ref{resmeas}(c) shows the standard deviations $\sigma_\mathrm{R}$ of the junction resistances across the wafers, scaled to the median value of each junction size. For wafers B and C, $\sigma_\mathrm{R}$ is below 2.5\% for large junctions with dimensions $>$ 1 $\mu$m. For wafer A the deviation is somewhat larger, $<$6\% for junctions down to the size of 1.7 $\mu$m and at a somewhat elevated level below this. In all wafers the deviation increases with the decreasing junction size which is readily understood to be due to the increasing uncertainty of the junction dimension as the smallest junctions have the dimensions comparable to or smaller than the linewidth reduction. Yet, for all three wafers the standard deviation for junctions down to the size of 0.4 $\mu$m is below 23\% across the wafer. Even the smallest junctions down to the size of approximately 0.2 $\mu$m still well follow the dimensional scaling as apparent in Fig. \ref{resmeas}(b) though the deviation increases up to the range of 13\% - 76\% depending on the wafer. Furthermore, even though the spread of the smallest junctions across the wafer is relatively large, it was observed that for all junction sizes the extremal resistance difference $\left(\max{R}-\min{R}\right)/\left(\max{R}+\min{R}\right)$ within a single chip on wafer A was below 10\% for a great majority of the chips. This is excluding the few individual junctions falling below the trend as apparent in Fig. \ref{resmeas}(b).

\begin{figure}[t!]
\begin{center}
\includegraphics[width=15cm]{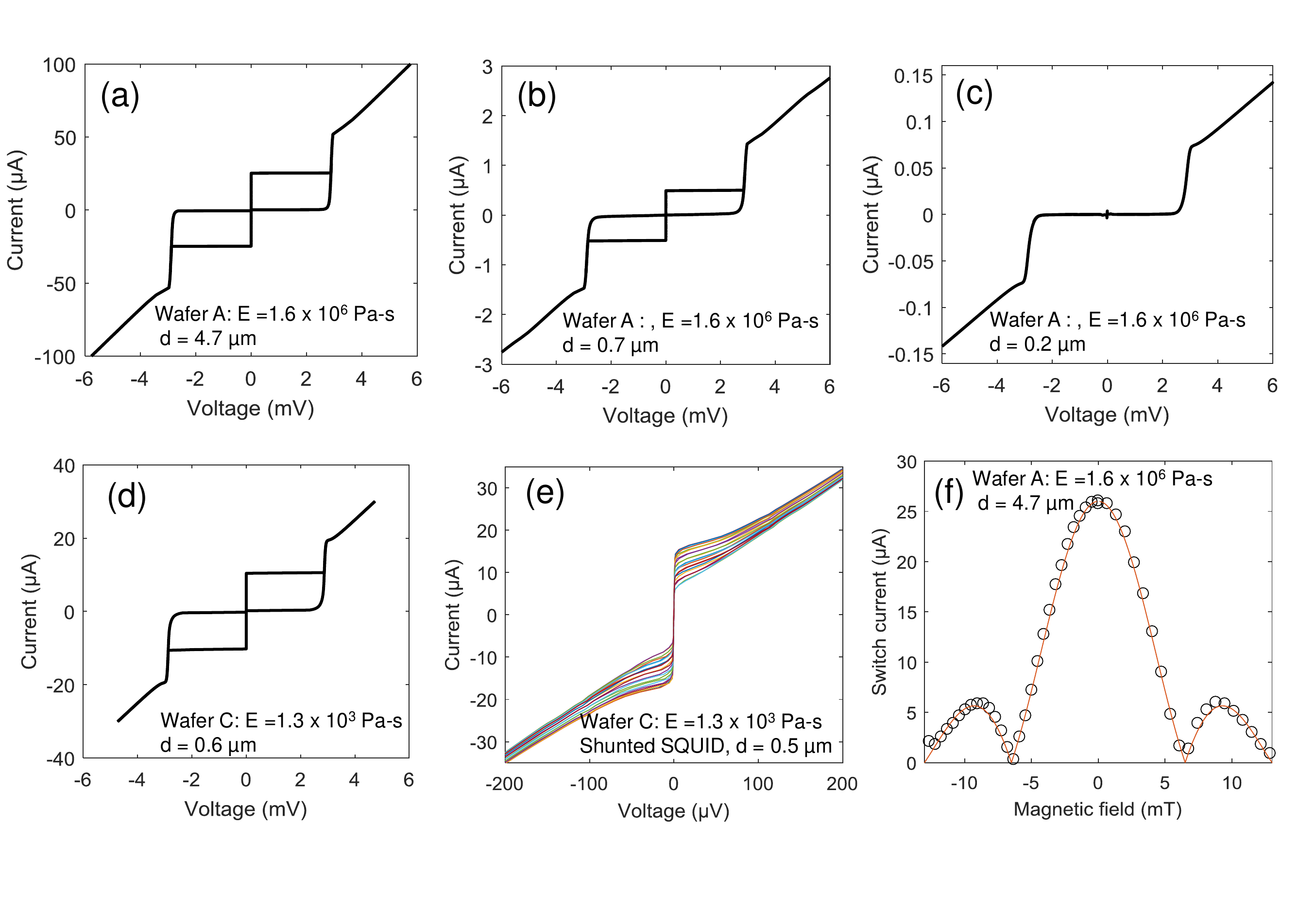}
\end{center}
 \caption{IV data from shunted and unshunted test structures from two different wafers differing by about 3 orders of magnitude in oxygen exposure used in the tunnel barrier formation. Wafer identifier and corresponding oxygen exposure $E$, as well as the junction dimension $d$, i.e. the side length of the square-shaped junction are shown in each frame. Frames (a)-(d) represent current-voltage characteristics of unshunted junctions. Frame (e) represents a family of current-voltage curves of a shunted dc SQUID at different flux operating points. In frame (f) the circles depict the magnetic field dependency of the switching current of an unshunted junction, and the solid line is a fit to the theory. The cryostat temperature is in the range of 9 - 20 mK for the data of frames (a)-(c) and (f), about 500 mK for the data of frame (d), and 4.2 K for the data of frame (e).}\label{IV}
\label{System}
\end{figure}

The tunnel junction resistivity $R_\mathrm{s}$ is plotted as the function of oxygen exposure in Fig. \ref{resmeas}(d). The power-law fit yields the critical exponent of 0.50$\pm$0.01. Assuming that the critical current $I_\mathrm{c}$ of the junction is inversely proportional to $R_\mathrm{s}$ this is in line with the empirical data predicting $J_\mathrm{c}\propto E^\alpha$ with $\alpha$= -0.4 \cite{exposure1} or $\alpha$ = -0.5 \cite{exposure2}. We have previously obtained an empirical prediction for the critical current density of Nb/AlO$_x$ junctions as $J_\mathrm{c} = 1.22 \textrm{ mV}/R_\mathrm{s}$ \cite{gro1}. The resistance data thus predicts the critical current densities of 0.1 kA/cm$^2$, 2.2 kA/cm$^2$ and 3.3 kA/cm$^2$ for the three wafers. This is verified by low-temperature measurements presented in Fig. \ref{IV} showing junction characteristics from wafers A and C. The quasiparticle branches of the IV curves of the unshunted junctions verify the basic functionality. The energy gap 2$\Delta$ is approximately 3.0 meV as expected for Nb. Furthermore, there are no excessive subgap leakage currents, though in our experiments the leakage was as limited by the setup.  For the junctions of the low-Jc wafer A (Figs. \ref{IV}(a-c)) the currents corresponding to the switching from the zero-voltage state are 25 $\mu$A, 0.50 $\mu$A, and 0.004 $\mu$A for the three junctions with square dimensions of 4.7 $\mu$m, 0.7 $\mu$m, and 0.2 $\mu$m, respectively. The two largest junctions have critical current values within 10\% as predicted from the room temperature data. Their $I_\mathrm{c} R_\mathrm{N}$ products are 1.4 mV and 1.1 mV, respectively, well in line with the niobium junctions reported elsewhere \cite{hyp1,jen1}. Here $R_\mathrm{N}$ is the asymptotic normal state resistance of the junction. The switching current of the smallest junction is strongly suppressed which is not surprising as the characteristic energy $hf_\mathrm{p}$ of the quantum fluctuations at the junction plasma frequency, where $h$ is the Planck constant and $f_\mathrm{p}$ the plasma frequency, is in this case several times higher than the Josephson coupling energy $\Phi_0 I_\mathrm{c}/(2\pi$) with $\Phi_0$ the flux quantum. In the unshunted junction of the high-$J_\mathrm{c}$ wafer C (Fig. \ref{IV}(d)) with $d$ = 0.6 $\mu$m  the switching occurs at 10.3 $\mu$A yielding the critical current density of 2.9 kA/cm$^2$, and $I_\mathrm{c} R_\mathrm{N}\approx$1.6 mV. Figure \ref{IV} (c) shows a family of IV curves for an externally shunted dc SQUID with a junction size of $d=$0.5 $\mu$m. The maximum zero-voltage current $2I_\mathrm{c} \approx$ 15 $\mu$A yields a critical current density of 3.0 kA/cm$^2$. Thus, also for wafer C the critical current densities are within 15\% as compared to the room temperature prediction, and the junction quality is basically as expected for Nb technology. Finally, Fig. 4(f) depicts the switching current of an unshunted junction in the external magnetic field $B$ aligned with the tunnel barrier. The dependence is of form $\sin{\pi\Phi}/\left(\pi\Phi_0\right)$ as expected with $\Phi \approx 2B\lambda_\mathrm{L} d$ the magnetic flux passing through the junction barrier. The fit yields the London penetration depth of $\lambda_\mathrm{L}\approx$ 34 nm. In Fig. 4, the  dc SQUID data was measured at the liquid He temperature of 4.2 K while other devices were measured at sub-Kelvin temperatures. In terms of observed critical currents the temperature scale is expected to have a rather minor effect. The effect for Nb at or below $T =$ 4.2 K is about 3\% as predicted from the temperature dependence $\propto\tanh{\left(\Delta/2k_\mathrm{B} T\right)}$ with $k_\mathrm{B}$ the Boltzmann constant \cite{amb1}.

\section{Conclusions}
We have presented the SWAPS structure which is effective in the precise definition of superconducting tunnel junctions. The junction characterization measurements yielded uniform junction data across the wafers indicating that the concept is functional. We verified successful lateral junction dimension scaling down to 0.2 $\mu$m, nearly a factor of two below the nominal lithography linewidth, and critical current density tunability in the range of 0.1 - 3 kA/cm$^2$ suggesting potential in various applications. On top of the junction process we have developed process versions optimized for SQUID magnetometers and Josephson parametric amplifiers, both of which have been up to date verified to produce functional devices. The SWAPS scheme differs from most other junction definition schemes  based on optical lithography in the sense that the dielectric materials typically inherent in junction definition methods \cite{hyp1,aist1,jen1,dan1} are removed from the device area. This is anticipated to be a benefit especially in sub-Kelvin applications utilizing the microwave band. In this sense our process resembles e-beam lithography and liftoff based junction definition schemes in which the junction is contacted over a smooth edge crossing \cite{pop1,wu1} . The SWAPS structure extends a similar approach to thicker layers and a wider range of lithography and etching techniques as verified in our case with niobium trilayers and optical lithography. Finally, we note that the SWAPS technique can also potentially be utilized in other tunnel junction applications beyond superconductive electronics. These include, for example, Coulomb blockade thermometry \cite{pek1} and magnetic junctions \cite{ota1}. In principle all devices where cross-line patterned functional junction can be used could benefit from SWAPS as the method is not limited to the Nb tri-layer junction of this work.

\section*{Acknowledgement}
We thank Harri Pohjonen for help with lithographic masks, and Paula Holmlund for assistance in sample preparation. This work has received funding from the Academy of Finland through grants 287768 and 284594, and through the European Union’s Horizon 2020 research and innovation programme under grant agreement No 686865.

\end{document}